\documentclass[12pt]{article}
\usepackage{amsmath,amssymb}

\usepackage{graphicx}
\usepackage{wrapfig}
\usepackage{tikz}
\usetikzlibrary{decorations.markings}
\usetikzlibrary{arrows}

\usepackage{hyperref}
\usepackage{cite}

\def\beq{\begin{eqnarray}}
\def\eeq{\end{eqnarray}}

\def\eps{\epsilon}

\newcommand{\Tr}{\,\mathrm{Tr}\,}            

\newcommand{\be}{\begin{equation}}
\newcommand{\ee}{\end{equation}}
\newcommand{\bea}{\begin{eqnarray}}
\newcommand{\eea}{\end{eqnarray}}
\newcommand{\bg}{\begin{gather}}

\newcommand{\bseq}{\begin{subequations}}
\newcommand{\eseq}{\end{subequations}}

\renewcommand{\ln}{\mathop{\rm ln}\nolimits}

\def\tr{\hbox{Tr}}

\def\be{\begin{eqnarray}}
\def\ee{\end{eqnarray}}
\def\lb{\label}

\begin{document}

\title{\textbf{Boundary effects in entanglement entropy}}

\vspace{2cm}
\author{ \textbf{  Cl\'ement Berthiere and  Sergey N. Solodukhin}} 
\date{}
\maketitle
\begin{center}
  \emph{  Laboratoire de Math\'ematiques et Physique Th\'eorique  CNRS-UMR
7350, }\\
  \emph{F\'ed\'eration Denis Poisson, Universit\'e Fran\c cois-Rabelais Tours,  }\\
  \emph{Parc de Grandmont, 37200 Tours, France}
\end{center}



\begin{abstract}
\noindent { We present a number of explicit calculations of Renyi and entanglement entropies in situations
where the entangling surface intersects the boundary in $d$-dimensional Minkowski spacetime. When the boundary is a single plane we compute the 
contribution to the entropy due to this intersection,
first in the case of the Neumann and Dirichlet boundary conditions, and then in the case of a generic Robin type
boundary condition. The flow in the boundary coupling between the Neumann and Dirichlet phases is analyzed in arbitrary dimension $d$ and is shown to be monotonic,  the peculiarity of $d=3$ case 
is noted.
We argue that the translational symmetry along the entangling surface is broken due to  the presence of the boundary
which reveals that  the entanglement is not homogeneous. In order to characterize this quantitatively, we introduce a density of entanglement entropy
and compute it explicitly. This quantity clearly indicates that the entanglement is maximal near the boundary.
We then consider the situation  where the boundary is composed of two parallel planes at a finite separation and compute the entanglement entropy as well as its
density in this case. The complete contribution to entanglement entropy due to the boundaries is shown not to depend on the distance between
the planes and is simply twice the entropy in the case of single plane boundary. Additionally, we find how the area law, the part in the entropy proportional to the area of entire
entangling surface, depends on the size of the separation between the two boundaries. The latter is shown to appear in the UV finite part of the entropy.
}
\end{abstract}

\vskip 1 cm
\noindent
\rule{7.7 cm}{.5 pt}\\
\noindent 
\noindent
\noindent ~~~ {\footnotesize e-mail:  clement.berthiere@lmpt.univ-tours.fr, Sergey.Solodukhin@lmpt.univ-tours.fr}


\newpage

\section{ Introduction}
Entanglement entropy is a useful tool which plays an important role in modern physics. 
First introduced \cite{EE} in order to explain the black hole entropy, it was later shown to be 
very efficient in measuring the quantum entanglement between sub-systems separated
by a surface. In infinite spacetime this surface is necessarily compact so that it divides the
spacetime into two complementary regions. The correlations present in the quantum system across the
entangling surface produce the non-trivial entropy which is essentially determined by the
geometry of the surface. The geometrical nature of entanglement entropy explains why it finds so many applications
in various fields of physics, from black holes and holography to integrable models and quantum computers
\cite{entanglement:reviews}.   For some recent progress in measuring entanglement entropy see \cite{exp}.

For conformal field theories, the entanglement entropy plays a special and important role since the
logarithmic terms in the entropy are related to the conformal anomalies, as suggested in \cite{Solodukhin:2008dh}.
In infinite spacetime,  the anomaly appears only
in even dimensions. In parallel, for compact entangling surfaces, only in even dimensions there appear
the logarithmic terms in the entropy.

Recently there has been some progress in understanding the conformal anomalies in the case where the spacetime is not infinite but has some boundaries,
\cite{Herzog:2015ioa}, \cite{Fursaev:2015wpa}, \cite{Solodukhin:2015eca}, \cite{Huang:2016rol} (for earlier works see \cite{Dowker:1989ue}).
It is interesting that in the presence of boundaries   the integrated anomaly is non-vanishing in odd spacetime dimensions, the relevant contribution being produced by the 
boundary terms only, \cite{Solodukhin:2015eca}. Thus, it becomes an interesting and urgent problem to understand the precise structure of 
the entropy for entangling  surface which intersects the boundary of a spacetime. In the holographic context, this and related problems were studied in
\cite{Takayanagi:2011zk}, \cite{Fujita:2011fp}, and on the field theory side  in \cite{Fursaev:2013mxa}.
The precise calculation for free fields of various spin in dimension $d=3$ has been done in \cite{Fursaev:2016inw} where it was shown that the logarithmic term
in the entropy in this case is proportional to the number of intersections the entangling surface has with the boundaries.
In higher dimensions it was  suggested that, unlike the case of compact closed surfaces,  the logarithmic terms in the entropy of a surface intersecting the boundary are present in any, odd and even, dimensions. 

The boundary phenomenon in entanglement entropy is certainly more general and is not restricted only to conformal field theories.
Yet, the explicit calculations for arbitrary boundaries and surfaces are technically complicated, if even possible.
Therefore, we find it instructive to first analyze the problem in some simple cases, where the spacetime is flat and the boundary is composed by a collection of planes.
In this paper we present a number of explicit calculations, for a free massive scalar field, of entanglement entropy in the case where the entangling surface is
a plane which crosses orthogonally the boundary. The main focus is made on the role of the boundary conditions. The latter can be viewed as some form of boundary interactions.
The general Robin type condition then interpolates between the Neumann condition in the weak coupling regime  and the Dirichlet condition in the strong coupling regime.
We study the respective behavior of entanglement entropy when the boundary coupling passes between these two regimes.

The paper is organized as follows. In Section 2  we  review the standard replica method that uses   the heat kernel and   the conical singularity technology.
We demonstrate how this method works for a simple case of infinite plane in infinite (without boundaries) Minkowski spacetime. This technology is then applied in Section 3 to the case of a single plane boundary with the Neumann (Dirichlet) boundary condition. The case of a general Robin type condition is considered in Section 4. We observe some inequalities for  the entropy for different boundary conditions in Section 5.  The monotonicity of the
entropy with respect to the boundary coupling is demonstrated in Section 6. Two parallel boundaries and the effects of the finite size are
considered in Section 7.  In Section 8 we introduce a notion of the  entanglement entropy density and calculate this quantity in all examples considered in the previous sections.  We conclude in Section 9.

\section{Replica method, heat kernel and entanglement entropy}
Before proceeding, we remind the
technical method very useful for calculation of entanglement
entropy. This method is known as {\it the replica method}. One first observes that $-\Tr \rho \ln
\rho=-(\alpha\partial_\alpha-1)\ln \Tr \rho^\alpha|_{\alpha=1}$. The
next observation is that  the density matrix obtained by tracing
over modes inside the surface $\Sigma$ is
$\ln\tr\rho^\alpha=-W[\alpha]$, where $W[\alpha]=-\ln Z(\alpha)$ and
$Z(\alpha)$ is the partition function of the field system in
question, considered on Euclidean space with a conical singularity at the
surface $\Sigma$. Thus one has that 
\be
S=(\alpha\partial_\alpha-1)W(\alpha)|_{\alpha=1}~~. 
\lb{SS} 
\ee
 One chooses the local
coordinate system $\{X^\mu=(\tau,x_i)\},$ where $\tau$ is the Euclidean time,
such that the surface $\Sigma$ is defined by the conditions
$\tau=0,\,x_1=0$ and $(x_2,..,x_d)$ are the coordinates on $\Sigma$.
In the subspace $(\tau,x_1)$ it is convenient to choose the polar
coordinate system $\tau=r\sin\phi$ and $x=r\cos\phi$ where
angular coordinate $\phi$ changes in the limits $0\leq \phi <
2\pi$. The conical space in question is then defined by making the
coordinate $\phi$ periodic with the period $2\pi\alpha$, where
$(1-\alpha)$ is very small.

In order to calculate the effective action $W(\alpha)$ we use the
heat kernel method. 
Consider a quantum bosonic field  described by a field operator
$\cal D$ so that $Z=\det^{-1/2}{\cal D}$. Then the effective
action is defined as 
\be W=-\frac{1}{2}\int_{\epsilon^2}^\infty
\frac{ds}{s}\Tr K
\lb{W}~, 
\ee 
where $\epsilon$ is an UV cut-off,
and is expressed by means of the trace of the heat kernel
$K(X,X',s)=\langle X|e^{-s{\cal D}}|X'\rangle$ satisfying the heat
kernel equation 
\be &&(\partial_s+{\cal
D})K(X,X',s)=0~,\nonumber \\
&&K(X,X',s=0)=\delta(X,X') ~.
\lb{K} 
\ee 
In the Lorentz invariant
case, the heat kernel $K(\phi,\phi',s)$ (where we skip the
coordinates other than the angle $\phi$) on regular flat space depends
on the difference $(\phi-\phi')$. The heat kernel $K_\alpha
(\phi,\phi',s)$ on space with a conical singularity is then
constructed from this quantity by applying the Sommerfeld formula
\cite{Sommerfeld}
\be 
K_\alpha(\phi,\phi',s)=K(\phi-\phi',s)
+\frac{\imath}{ 4\pi\alpha}\int_\Gamma \cot \frac{w}{ 2\alpha}
K(\phi-\phi'+w,s)dw~. 
\lb{Sommerfeld} 
\ee 
The contour $\Gamma$
consists of two vertical lines, going from $(-\pi+i\infty )$
to $(-\pi-i\infty )$ and from $(\pi-i\infty )$ to
$(\pi+i \infty )$, and intersecting the real axis between the
poles of $\cot \frac{w}{ 2\alpha}$: $-2\pi\alpha$, $0$ and $0,$
$+2\pi\alpha$, respectively. For $\alpha=1$ the integrand in
(\ref{Sommerfeld}) is a $2\pi$-periodic function and the
contributions of these two vertical lines cancel each other. Thus,
for a small angle deficit the contribution of the integral in
(\ref{Sommerfeld}) is proportional to $(1-\alpha)$.

In $d$-dimensional spacetime,
for a massive scalar field described by the operator \mbox{${\cal D}=\nabla^2+m^2$},
$\nabla^2=\partial^2_\tau+\sum_{i=1}^{d-1}\partial^2_{i}$, where
$\tau$ is the Euclidean time, the heat kernel is known explicitly,
\be
K(\tau,\tau',x,x',s)=\frac{e^{-m^2s}}{(4\pi s)^{d/2}}e^{-\frac{1}{
4s}[(\tau-\tau')^2+\sum_i (x_i-x'_i)^2]} \,.
\lb{Kd} 
\ee 
We take a
$(d-2)$-surface $\Sigma$ to be the infinite plane defined by equations
$x_1=0, \ \tau=0$ so that $(x_2,x_3,..,x_d)$ are coordinates on
$\Sigma$.  
In the polar coordinate system $\tau=r\sin\phi$ and
$x_1=r\cos\phi$ we have for two points $(r,\phi)$ and $(r,\phi')$
that $(\tau-\tau')^2+(x_1-x'_1)^2=4r^2\sin^2(\frac{\phi-\phi'}{
2})$. The trace is defined as $\Tr K_\alpha=\int
d^{d-2}x_i\int_0^\infty dr\ r\ \int_0^{2\pi\alpha}d\phi
K_\alpha(\phi=\phi',r'=r,x_i=x'_i,s)$. For the contour integral
over $\Gamma$ one finds  (see  \cite{coneQFT}) 
\be 
C_2(\alpha)\equiv \frac{i}{8\pi\alpha}\int_\Gamma \cot\frac{w}{2\alpha}\ \frac{dw}{
\sin^2\frac{w}{2}}=\frac{1}{6\alpha^2}(1-\alpha^2)~. 
\lb{C} 
\ee
Thus one obtains for the trace of the heat kernel
\be \Tr K_\alpha=\frac{e^{-m^2s}}{(4\pi s)^{d/2}}\left(\alpha V+s \
2\pi\alpha C_2(\alpha)A(\Sigma)\right)~, \lb{TrK} 
\ee 
where
$V=\int d\tau d^{d-1}x$ is the volume of spacetime and
$A(\Sigma)=\int d^{d-2}x$ is the area of the surface $\Sigma$.
The entanglement entropy is then easily obtained,
\be
S_d(\Sigma) = \frac{ A(\Sigma)}{12(4\pi)^{(d-2)/2}} \int_{\epsilon^2}^\infty ds\, \frac{e^{-sm^2}}{s^{d/2}}\,.
\lb{Sd}
\ee
We stress that this is the entropy for an infinite plane $\Sigma$ in infinite (without boundaries) Minkowski spacetime. 
For the UV divergent part of the entropy we have
\be
S_d(\Sigma) = \frac{ A(\Sigma)}{6(4\pi)^{(d-2)/2}} \sum_{k=0}^{[\frac{d-2}{2}]}\frac{(-1)^k m^{2k}\eps^{2k+2-d}}{k!(d-2k-2)} \,.
\lb{SSd}
\ee
In even dimension $d$ the term with $k=(d-2)/2$ becomes a logarithm. 

The R\'enyi entropy is defined by the formula
\be
S(n)=\frac{\ln\Tr \rho^n-\ln\Tr \rho}{1-n}\, .
\lb{Renyi}
\ee
Thus, in order to compute this entropy one needs to keep finite $\alpha=n$ in  (\ref{C}) and (\ref{TrK}).  One finds that
in our example of infinite plane in Minkowski spacetime   the Renyi entropy is simply proportional to the entanglement entropy,
\be
S(n)=\frac{1}{2}(1+n^{-1})S_{\rm ent}\, .
\lb{SS}
\ee
In all examples considered in this paper we have a similar relation between the two entropies. In what follows we thus keep our focus on computing the entanglement entropy.

\section{Single plane boundary: Neumann and Dirichlet  boundary conditions}
Consider $d$-dimensional flat spacetime with coordinates $X^\mu= (\tau, x, y, z_i, i = 1, .., d-3)$ and a plane boundary at $y=0$. 
We define the entangling surface $\Sigma$ by the equations: $\tau=0,~ x=0$. It crosses the boundary $\partial M_d$ orthogonally, the intersection is
$(d-3)$-surface $P$ with coordinates $\{z_i, ~ i=1,.., (d-3)\}$.
We impose Neumann or Dirichlet boundary condition at $y=0$,
\be
\partial_{y}K^{(N)}\Big|_{y=0} = 0\,, \qquad  {\rm or} \,  \qquad  K^{(D)}\Big|_{y=0} = 0 \,.
\lb{K-N}
\ee
The solution to the heat kernel equation (\ref{K}) with this boundary condition is constructed from the heat kernel (\ref{Kd}) on infinite spacetime as follows
\be
&&K^{N(D)}(s,X, X')= \nonumber \\
&&\frac{e^{-m^2s}}{(4\pi s)^{d/2}}\left(e^{-\frac{1}{
4s}[(\tau-\tau')^2+(x-x')^2+(y-y')^2+(z-z')^2]} \pm e^{-\frac{1}{
4s}[(\tau-\tau')^2+(x-x')^2+(y+y')^2+ (z-z')^2]} \right) \, ,
\lb{KND0}
\ee
where the plus (minus) corresponds to Neumann (Dirichlet) condition. 
Then  we are supposed to go through the same steps as before.
Taking the  trace, i.e. identifying $\varphi'=\varphi+w$ and $y=y', \, z=z'$, and 
taking the contour integral over $w$ and the integration over $\varphi$, $y$ and  $z$, we find
\be
\Tr K^{N(D)}_\alpha(s)\;=\; \Tr K_\alpha (s)\pm \frac{\alpha(\alpha^{-2}-1)}{12(4\pi)^{(d-2)/2}} \frac{e^{-sm^2}}{s^{(d-2)/2}}A(P)\int_0^\infty dy \,e^{-y^2/s} \, ,
\lb{TrKN}
\ee
where  the first term is the same as in infinite (without boundaries) spacetime and  $A(P)=\int d^{d-3}z$ is the area of $P$. We then use that $\int_0^\infty dy\, e^{-y^2/s}=\frac{\sqrt{\pi s}}{2}$.
Applying the replica trick and computing the integration over proper time $s$ we arrive at the following form of the entanglement entropy,
\be
&&S^{N(D)}_d(\Sigma) \;=\; S_d(\Sigma)  \pm S_d(P)\, , \\
&&S_d(P)\;=\;\frac{A(P)}{48(4\pi)^{(d-3)/2}} \int_{\epsilon^2}^\infty ds\, \frac{e^{-sm^2}}{s^{(d-1)/2}}\, .\nonumber
\lb{SND}
\ee
Here $S_d(\Sigma)$ is the entropy in infinite spacetime, defined in (\ref{Sd}), and $S_d(P)$ is the part of the entropy which is entirely due to the intersection $P$ of the entangling surface 
$\Sigma$ and the boundary $\partial M_d$.
For the UV divergent part of this entropy one finds
\be
S_d(P)= \frac{ A(P)}{24(4\pi)^{(d-3)/2}\epsilon^{d-3}} \sum_{k=0}^{[\frac{d-3}{2}]}\frac{(-1)^k m^{2k}\eps^{2k}}{k!(d-2k-3)} \, .
\lb{SdP}
\ee
In particular, for $d=3,\,4$ dimensions we find
\be
S_3(P)=-\frac{1}{24}\ln({\epsilon m})\, , \ \ 
S_4(P)=\frac{A(P)}{48\epsilon\sqrt{\pi}}\, .
\lb{d=3,4}
\ee
The $d=3$ case was already considered in \cite{Fursaev:2016inw}.
We see from (\ref{SdP}) that there  appears a  logarithmic term in $S_d(P)$ if spacetime dimension $d$ is odd.
Thus, there always appears a logarithm in the complete entanglement entropy: either due to  $S_d(\Sigma)$ in even dimension $d$
or due to $S_d(P)$ in odd dimension $d$.

\section{Single plane boundary: Robin boundary condition }
We now generalize the above analysis and consider a more general boundary condition of the Robin type,
\be
(\partial_y-h)K^{(h)}\Big|_{y=0} = 0 \, , 
\lb{robin}
\ee
where $h$ is the boundary coupling constant. Value $h=0$ corresponds to the Neumann  boundary condition while the limit $h\rightarrow +\infty$ corresponds to the
Dirichlet boundary condition. The corresponding solution to the heat kernel equation (\ref{K}) takes the form (see  \cite{Solodukhin:1998ec}),
\be
K^{(h)}(s, y,y') &=& K^{(N)}(s, y,y') -2h\, e^{h(y+y')}\int_{y+y'}^\infty d\sigma\,e^{-h\sigma}K(s, \sigma)\, ,\qquad
\lb{KRobin}
\ee
where $y$ is the coordinate orthogonal to the boundary and  we skip all other coordinates. The trace of this heat kernel considered on spacetime with a conical singularity
reads
\be
\Tr K_\alpha^{(h)}(s)=\Tr K^{(N)}_\alpha (s)-A(P)\alpha(\alpha^{-2}-1) \frac{ e^{-s(m^2-h^2)}}{24(4\pi s)^{(d-3)/2}} (e^{-h^2s}+\Phi(h\sqrt{s})-1)\, ,
\lb{TrKR}
\ee
where $\Phi(x)=\frac{2}{\sqrt{\pi}}\int_0^x e^{-t^2}dx$ is the error function. Respectively we find for the entanglement entropy,
\be
S^{(h)}_d(\Sigma) &=& S^{(N)}_d(\Sigma)  -  \frac{A(P)}{24(4\pi)^{(d-3)/2}} \int_{\epsilon^2}^\infty ds\, \frac{e^{-sm^2}}{s^{(d-1)/2}}\left(1+e^{h^2s}(\Phi(h\sqrt{s}) -1) \right)\, .\qquad
\lb{Sh}
\ee
For positive boundary coupling $h>0$ and in the limit of large $s$ the function which appears under the integral in (\ref{Sh}) behaves as
\be
F(h\sqrt{s})\equiv\left(1+e^{h^2s}(\Phi(h\sqrt{s}) -1) \right)=1-\frac{1}{\sqrt{\pi s}~h}+\mathcal{O}(s^{-3/2})\, ,\quad h>0 \, .
\lb{limit}
\ee
Therefore, the integral in (\ref{Sh}) converges in the upper limit in dimension $d>3$, even in the massless case ($m=0$) if the coupling $h$ is positive.
On the other hand, for negative $h<0$ one has
\be
F(h\sqrt{s})=-2e^{sh^2}+1 + \mathcal{O}(s^{-1/2})\, , \quad h<0 \, 
\lb{limit2}
\ee
and the integral in (\ref{Sh}) converges in the upper limit only if the mass is sufficiently large,  $m^2 > h^2$. 

On the other hand, for small $s$ we find
\be
F(h\sqrt{s})=\frac{2h}{\sqrt{\pi}}\sqrt{s}+\mathcal{O}(s)\, .
\lb{limit3}
\ee
Therefore, we note that in dimension $d\geq 4$ the integral in (\ref{Sh}) is divergent when the lower limit is taken to zero and thus the regularization with $\epsilon$ is 
needed. This is of course the usual UV divergence.  However,
in dimension $d=3$ the integral in (\ref{Sh}) has a regular limit if $\epsilon$ is taken to zero. Thus, for any finite $h$ the integral in (\ref{Sh}) is UV finite.
The integration can be performed explicitly in dimension $d=3$ and one finds
\be
S^{(h)}_3(\Sigma) = S^{(N)}_3(\Sigma) -  \frac{1}{12}\ln\left(1+\frac{h}{m}\right)\, , \hspace{1cm} (m>-h)\, .
\lb{Sd=3}
\ee
It is interesting that this is the exact result.
We see that in this case the boundary coupling appears only in the UV finite term in the entropy. We notice that the entropy (\ref{Sd=3}) is divergent if $m+h\rightarrow 0$. This is a IR 
divergence: the integral in (\ref{Sh}) diverges  in the upper limit if  $h$ is negative  and $h<-m$. 

In higher dimensions the integration can be done in a  form of an expansion in powers of $h$,
\be
S^{(h)}_4(\Sigma) &=& S^{(N)}_4(\Sigma) + \frac{A(P) }{12\pi}h\ln\epsilon +\mathcal{O}(h^2)  \, ,  \lb{Sd456}\\
S^{(h)}_5(\Sigma) &=& S^{(N)}_5(\Sigma)- \frac{A(P)}{48\pi} \left( \frac{2}{\sqrt{\pi}}\frac{h}{\epsilon} + h^2\ln\epsilon +\mathcal{O}(h^3) \right)   \, , \nonumber \\
S^{(h)}_6(\Sigma) &=& S^{(N)}_6(\Sigma) - \frac{A(P)}{96\pi^{2}} \left(\frac{h}{\epsilon^2}  + 2hm^2\ln\epsilon - \frac{h^2}{\epsilon}\sqrt{\pi}  - \frac{4}{3}h^3\ln\epsilon +\mathcal{O}(h^4)\right)   \, . \nonumber 
\ee
More generally, we find the expansion in arbitrary dimension $d$,
\be
S^{(h)}_d(\Sigma) &=&S^{(N)}_d(\Sigma)  - \frac{A(P)}{12(4\pi)^{(d-3)/2}} \sum_{k=0}^\infty \left[\frac{a_k}{(d-4-2k)\epsilon^{d-4-2k}} + \frac{b_k}{(d-5-2k)\epsilon^{d-5-2k}} \right] \, ,
\lb{x}
\ee 
where
\be
a_k &=& \frac{2h^{2k+1}}{\sqrt{\pi}}\frac{(-1)^k}{k!(2k+1)}\,_2F_1\left(-k-\frac{1}{2}, -k, -k+\frac{1}{2}, 1-\frac{m^2}{h^2} \right)  \, , \nonumber\\
b_k &=& \frac{(-1)^{k+1}}{(k+1)!}\left((m^2)^{k+1}-(m^2- h^2)^{k+1} \right)  \, .
\lb{ab}
\ee
To leading order in $h$ we  find in any dimension $d$,
\be
S^{(h)}_d(\Sigma) &=&S^{(N)}_d(\Sigma)  - \frac{h\, A(P)}{6(4\pi)^{(d-2)/2}(d-4)\epsilon^{d-4}}  \, .
\ee 
In dimension $d=4$ the power law is replaced by a logarithm as in (\ref{Sd456}).

The integral  (\ref{Sh}) is divergent in the upper limit  if $h<-m$. Therefore the entropy shows a divergence when $(h+m)$ goes to zero.
This is a IR divergence. In dimension $d=3$ this divergence is logarithmic. In higher dimension $d>3$ the divergence is milder. The entropy takes a finite value if $h=-m$. However, the derivatives of sufficiently high order diverge there
\be
&&S^{(h)}_d(\Sigma)\sim  (m^2-h^2)^{\frac{d-3}{2}}\, , \ \ d \   {\rm even} \nonumber \\
&&S^{(h)}_d(\Sigma)\sim  (m^2-h^2)^{\frac{d-3}{2}}\ln (m^2-h^2)\, , \ \ d \   {\rm odd}
\lb{ed}
\ee
so that the entropy is not an analytic function of $h$ at the point $h=-m$. This may signal for some type of a phase transition. We, however,
do not elaborate on this idea here.

The other useful forms of  (\ref{Sh}) are
\be
S^{(h)}_d(\Sigma) &=& S^{(D)}_d(\Sigma)  -  \frac{A(P)}{24(4\pi)^{(d-3)/2}} \int_{\epsilon^2}^\infty ds\, \frac{e^{-sm^2}}{s^{(d-1)/2}}e^{sh^2}\left(\Phi(h\sqrt{s}) -1 \right)\, \qquad
\lb{Sh2}
\ee
that compares the Robin entropy with the entropy in the case of the Dirichlet boundary condition, and
\be
S^{(h)}_d(\Sigma)& =& S_d(\Sigma) +S_d^{(h)}(P) \, , \nonumber \\ 
S_d^{(h)}(P) &=&-  \frac{A(P)}{24(4\pi)^{(d-3)/2}} \int_{\epsilon^2}^\infty ds\, \frac{e^{-sm^2}}{s^{(d-1)/2}}\left(\frac{1}{2}+e^{sh^2}(\Phi(h\sqrt{s}) -1) \right)\,, 
\lb{Sh3}
\ee
that compares it with the entropy in the case of infinite (without boundaries) spacetime. This equation generalizes (\ref{SND}) for arbitrary boundary coupling $h$.

\section{Some inequalities}

Here we formulate some inequalities relating the entropies for various boundary conditions.
The first obvious inequality follows from equation (\ref{SND}). Indeed, it simply  indicates that the entropy for a field with
the Neumann boundary condition is strictly larger than that for the Dirichlet boundary condition,
\be
S_d^{(N)}(\Sigma)> S_d^{(D)}(\Sigma)\, .
\lb{s1}
\ee
Including the entropy computed for a plane of the same area in infinite spacetime, we have
\be
 S_d^{(D)}(\Sigma)< S_d(\Sigma) < S_d^{(N)}(\Sigma)\, .
\lb{s2}
\ee
The other inequalities come from the comparison with the entropy for the Robin boundary condition. 
Comparing the entropy for the Neumann and the Robin boundary conditions we use equation (\ref{Sh}).
The function $F(h\sqrt{s})$, introduced in (\ref{limit}), that appears  in the integral in (\ref{Sh}) is positive
for positive values of $h$ and negative for negative values,
\be
&&F(h\sqrt{s})>0 \, ,  \;\; h>0\,,\nonumber \\
&&F(h\sqrt{s})<0 \, , \;\; h<0\, . \qquad
\lb{s30}
\ee
On the other hand, the comparison with the entropy for the Dirichlet boundary condition uses equation (\ref{Sh2}). 
The function that appears in the integral in this case is negative for any (positive or negative) values of $h$,
\be
e^{h^2s}(\Phi(h\sqrt{s})-1) <0\, ,  \ \forall h\, .
\lb{s4}
\ee
Using  (\ref{s3}) and (\ref{s4}) we conclude that for positive values of $h$,
\be
S_d^{(D)}(\Sigma)< S_d^{(h)}(\Sigma) < S_d^{(N)}(\Sigma)\, , \;\  h>0 \,,
\lb{s5}
\ee
while for negative values of $h$,
\be
S^{(h)}_d(\Sigma)> S_d^{(N)}(\Sigma)> S_d^{(D)}(\Sigma)\, , \;\  h<0\,.
\lb{s6}
\ee
Thus, increasing the negative values of $h$ one makes the entanglement entropy larger than it is for the Neumann boundary condition.
However, one cannot make $h$ as negative as   one wants since, as we have shown, the integral in (\ref{Sh})
is not convergent for large $s$ if $h<-m$. 

On the other hand, for positive $h>0$, increasing the value of $h$ to infinity one arrives at the entanglement entropy for the Dirichlet boundary condition.
Indeed, using that 
\be
e^{h^2s}(\Phi(h\sqrt{s})-1)=-\frac{1}{\sqrt{\pi s} h}+\mathcal{O}\Big(\frac{1}{h^3}\Big)\, , \;\; h>0\,,
\lb{s7}
\ee
one finds from equation (\ref{Sh2}) that
\be
S^{(h)}_d(\Sigma)=S^{(D)}_d(\Sigma)+\frac{1}{h}\frac{A(P)}{12(4\pi)^{(d-2)/2}} \int_{\epsilon^2}^\infty ds\, \frac{e^{-sm^2}}{s^{d/2}}+\mathcal{O}\Big(\frac{1}{h^3}\Big)\, .
\lb{s8}
\ee
This relation indicates that in the limit $h\rightarrow +\infty$ the Robin entropy approaches the Dirichlet entropy,
\be
\lim_{h\rightarrow +\infty}S^{(h)}_d(\Sigma)= S^{(D)}_d(\Sigma)\, .
\lb{s9}
\ee
We stress that this limit is valid only if $h\epsilon\rightarrow \infty$ so that $1/h$ should be smaller than the UV cut-off.
The case of $d=3$ is special. In this case, the integral in (\ref{Sh2}) goes from $0$ to $\infty$ so that one necessarily includes the integration over small values of $s$.
Therefore, the approximation (\ref{s7}) cannot be justified for all values of $s$. In fact, the integration over $s$ can be performed explicitly. The result  (\ref{Sd=3})
of this integration shows that in this case the limit $h\rightarrow +\infty$ is divergent and the Robin entropy does not approach  the Dirichlet entropy.
We stress once again that this is a peculiarity  of three dimensions. Taking this observation it seems that the claim made in \cite{Jensen:2015swa} that 
in $d=3$ CFT the RG flow which starts in the Neumann phase should end in the Dirichlet phase should probably be taken with some caution.

\section{Monotonicity of flow  between Neumann and Dirichlet phases}

Above we have shown that, in dimension $d>3$, varying the boundary coupling $h$ from zero to plus infinity, the Robin entropy changes from the Neumann entropy to 
the Dirichlet entropy. An interesting question is whether this evolution of the entropy is monotonic? The answer to this question is affirmative as we now show.
Indeed, the derivative with respect to $h$ of the Robin entropy (\ref{Sh}) 
\be
\partial_h S^{(h)}_d(\Sigma)=-  \frac{A(P)}{24(4\pi)^{(d-3)/2}} \int_{\epsilon^2}^\infty ds\, \frac{e^{-sm^2}}{s^{(d-1)/2}}\partial_h F(h\sqrt{s})\;<\;0
\lb{m1}
\ee
is negative as follows form the fact that
\be
\partial_h F(h\sqrt{s})=2hs e^{h^2s}\left(\Phi(h\sqrt{s})-1+\frac{2}{\sqrt{\pi s}h}e^{-h^2s}\right)>0\, , \quad  h>0\,,
\lb{m2}
\ee
is positive for positive values of $h$. Thus, the  entropy is monotonically decreasing provided one changes the boundary coupling $h$ from zero to $+\infty$.
It goes from the Neumann entropy for $h=0$ to the Dirichlet entropy for $h=+\infty$.

This demonstration is also valid in dimension $d=3$. In fact, the monotonicity in this case can be seen directly from the exact formula (\ref{Sd=3}). However, in the limit $h\rightarrow +\infty$
it  does not approach the Dirichlet entropy. This is consistent with the discussion we made above.

\section{Two parallel plane boundaries}
We now want to analyze whether the boundary part in the entanglement entropy is affected by the finite size of the system.
We start with a simple case of two parallel plane boundaries, at $y=0$ and $y=L$. At each boundary one may impose either Neumann or Dirichlet
boundary condition so that we have three cases to consider
\begin{alignat}{4}
&{\rm Neumann-Neumann}\; &:& \qquad  \partial_y K^{NN}\Big|_{y=0}\, &=&\;\; \partial_y K^{NN}\Big|_{y=L}\, &=& \;\;0\, ,\quad \lb{NN}  \\
&{\rm Dirichlet-Dirichlet}&:& \qquad  K^{DD}\Big|_{y=0} &=&\;\; K^{DD}\Big|_{y=L} &=& \;\;0\, , \lb{DD}  \\
&{\rm Neumann-Dirichlet}&:& \qquad  K^{ND}\Big|_{y=0} &=&\;\; \partial_y K^{ND}\Big|_{y=L}&=& \;\;0\, .
\lb{ND}
\end{alignat}\,

\subsection{Neumann-Neumann (Dirichlet-Dirichlet) boundary conditions}
The explicit form for the corresponding heat kernel is
\be
K^{NN(DD)}(s,  y, y') = \sum_{k\in\mathbb{Z}} K(s, y+2Lk,y') \pm K(s,2Lk-y,y') \, ,
\lb{KND}
\ee
where the plus (minus) corresponds to Neumann (Dirichlet) condition and we keep only the dependence on coordinate $y$ orthogonal to the boundaries.
As before, we define the entangling surface $\Sigma$ by equations: $\tau=0$, $x=0$.
Repeating the conical space construction for this heat kernel we arrive at the following trace
\be
\Tr K^{NN(DD)}_\alpha(s) &=& \alpha\Tr K^{NN(DD)}_{\alpha=1}(s)\nonumber \\
&& \hspace{-2cm}+\frac{\alpha(\alpha^{-2}-1)}{12(4\pi)^{(d-2)/2}} \frac{s\,e^{-sm^2}}{s^{d/2}}\left(A(\Sigma)\sum_{k\in\mathbb{Z}}e^{-\frac{L^2}{s}k^2} \pm \frac{1}{2}A(P)\int_0^\infty dy \sum_{k\in\mathbb{Z}}e^{-\frac{(y-Lk)^2}{s}} \right)\, ,
\ee
where $P$ is the intersection of the entangling surface $\Sigma$ with both boundaries, so that it has two disconnected components, at each of the boundary.
Respectively we find for the entanglement entropy 
\be
S^{NN(DD)}_d(\Sigma) &=& S_d(\Sigma, L)\pm S_d(P)\, , \\
S_d(\Sigma, L) &=& \frac{A(\Sigma)}{12(4\pi)^{(d-2)/2}} \int_{\epsilon^2}^\infty ds\, \frac{e^{-sm^2}}{s^{d/2}}\sum_{k\in\mathbb{Z}}e^{-\frac{L^2}{s}k^2}\, , \nonumber \\
S_d(P) &=& \frac{A(P)}{24(4\pi)^{(d-2)/2}} \int_{\epsilon^2}^\infty ds\, \frac{e^{-sm^2}}{s^{d/2}}\int_{0}^{L}dy \sum_{k\in\mathbb{Z}}e^{-\frac{(y-Lk)^2}{s}}\, .\nonumber
\lb{SND2tp}
\ee 
The integration over $y$ can be performed explicitly,
\be
\int_0^\infty dy\, e^{-\frac{(y-Lk)^2}{s}}=\frac{\sqrt{\pi s}}{2}\left(\Phi \Big(\frac{Lk}{\sqrt{s}}\Big)-\Phi\Big(\frac{L(k-1)}{\sqrt{s}}\Big)\right)\, .
\lb{int}
\ee
The sum over images then will give us
\be
\sum_{k\in\mathbb{Z}}\Phi\left(\frac{Lk}{\sqrt{s}}\right)-\Phi\left(\frac{L(k-1)}{\sqrt{s}}\right) = 2 \, .
\lb{sum}
\ee
Remarkably, this result does not depend on the size $L$. We conclude that the part in the entropy that is due to the intersection $P$ of the entangling surface with the boundary
is not sensitive to the finite size $L$. The whole effect of the presence of the second boundary is that this part in the entropy simply doubles,
\be
S_d(P)=\frac{A(P)}{48(4\pi)^{(d-3)/2}} \int_{\epsilon^2}^\infty ds\, \frac{e^{-sm^2}}{s^{(d-1)/2}}\, ,
\lb{SP}
\ee
so that the entropy is proportional to the complete area of the disjoint components of the intersection $P$.  In dimension $d=3$, $A(P)=2$ is the number of intersections of the line
$\Sigma$ with the two boundaries.

The size $L$, however, will appear in the area law, the part proportional to the area of the entire surface $\Sigma$. In fact, this is the UV finite part of the entropy that
will depend on $L$. Indeed,  in  the sum over images the term with $k=0$ will produce the UV divergence already analyzed above and the terms with $k\neq 0$ will give us a UV finite
contribution. In order to identify this contribution we may interchange the order of the   integration over $s$ and summation over $k$. The integration (for $k\neq 0$) then gives us
\be
\int_0^\infty \frac{ds}{s^{d/2}}e^{-sm^2} e^{-\frac{L^2k^2}{s}} =
\begin{cases}\displaystyle
2\Big(\frac{m}{Lk}\Big)^{\frac{d-2}{2}}K_{\frac{d-2}{2}}(2mLk)\,, &\quad m>0\,, \vspace{5pt}\\ 
\displaystyle \frac{2}{d-2}\frac{\Gamma(d/2)}{(Lk)^{d-2}}\,,  & \quad m=0\, .
 \end{cases} 
 \lb{int2}
 \ee
Thus we find 
\be
S_d(\Sigma, L) &=& \frac{ A(\Sigma)}{12(4\pi)^{(d-2)/2}} \left(\int_{\epsilon^2}^\infty ds\, \frac{e^{-sm^2}}{s^{d/2}}+   {\cal S}(L,m)      \right)\, , \\
{\cal S}_d(L,m) &=& 
\begin{cases}\displaystyle
4\sum_{k=1}^\infty\Big(\frac{m}{Lk}\big)^{\frac{d-2}{2}}K_{\frac{d-2}{2}}(2mLk)\, , &\quad m>0\, , \\
  \displaystyle \frac{4}{d-2}\frac{\Gamma(d/2)}{L^{d-2}}\sum_{k=1}^\infty\frac{1}{k^{d-2}}\, ,& \quad m=0\, .
\end{cases}  
 \lb{kk}
\ee
Some particular cases are worth mentioning. 

\medskip

\noindent {\bf 1.} In the massless case ($m=0$) in dimension $d>3$ one has
\be
{\cal S}_d(L, m=0)=\frac{4}{d-2}\frac{\Gamma(\frac{d}{2})\zeta(d-2)}{L^{d-2}}\,,
\lb{SL}
\ee
so that it decays by a power law. For $d=3$ the zeta-function in (\ref{SL}) diverges. This is yet another manifestation of the IR divergence in $d=3$ dimensions that we have already discussed.

\medskip

\noindent {\bf 2.}  In dimension $d=3$ the integral (\ref{int2})  produces elementary function,
\be
\int_0^\infty \frac{ds}{s^{3/2}}e^{-sm^2} e^{-\frac{L^2k^2}{s}} =\frac{\sqrt{\pi}}{Lk}e^{-2mLk}\, ,
\lb{int3}
\ee
so that the sum over $k$ in (\ref{kk}) can be easily evaluated and we find that
\be
{\cal S}_{d=3}(L,m)=-2\frac{\sqrt{\pi}}{L}\ln(1-e^{-2mL}) \, .
\lb{s3}
\ee
We see that it decays exponentially for large $L$ and approaches a logarithm for small $L$,
\be
{\cal S}_{d=3}(L,m)&\simeq& 2\frac{\sqrt{\pi}}{L}e^{-2mL} \, , \  \ Lm\gg 1\nonumber \\
&\simeq& 2\frac{\sqrt{\pi}}{L}\ln(1/2mL) \, , \  \  Lm\ll 1
\lb{sd3}
\ee
Similarly, one can analyze the massless limit in (\ref{s3}). In this limit   there exists the IR divergence
we have already discussed. Therefore, a IR regulator should be kept. We find that in this limit in the UV  finite part in the entropy
there appears a new logarithmic term,
\be
S^{\rm fin}_{d=3}=\frac{1}{12} \ln \frac{1}{L}\, ,
\lb{14'}
\ee
where we used that  the area $A(\Sigma)=L\times  A(P)/2$ and that $A(P)=2$ in dimension $d=3$. 
We see that this term is in fact not determined by the area of surface $\Sigma$.
It is due to a combination of two factors: the intersection of entangling surface with the boundary  and the finite size $L$ of the system.
 The logarithmic term (\ref{14'}) resembles the entanglement entropy in two dimensions.
It would be interesting to understand better the origin of this logarithmic term. Since in the massless case the theory becomes conformal,
the logarithmic term (\ref{14'}) may be related to conformal  symmetry.

\noindent {\bf 3.} In dimension $d=5$, the sum over $k$ in (\ref{kk}) for $m>0$ gives
\be
{\cal S}_{d=5}(L,m)&=& \frac{\sqrt{\pi}}{L^3}\,{\rm Li}_3(e^{-2mL}) + 2\sqrt{\pi}\frac{m}{L^2}\,{\rm Li}_2(e^{-2mL}) \,,
\ee 
where ${\rm Li}_n(x)$ is the polylogarithmic function. The asymptotics are given below for any $d>3$.

\noindent {\bf 4.} In dimension $d>3$ we have
\be
{\cal S}_{d>3}(L,m)&\simeq& 2\sqrt{\pi}\,\frac{m^{\frac{d-3}{2}}}{L^{\frac{d-1}{2}}}e^{-2mL} \, , \quad Lm\gg 1\,,\\
&\simeq&\frac{4}{d-2}\frac{\Gamma(\frac{d}{2})\zeta(d-2)}{L^{d-2}}\,,  \;\; Lm\ll 1\,.
\lb{sdd}
\ee

\medskip

We see that for the boundary conditions of the same type (NN or DD) the UV finite part in the area law (\ref{kk}) is a positive
quantity.

\subsection{Mixed boundary conditions}
Now we impose the mixed boundary conditions (\ref{ND}). 
The explicit form for the corresponding heat kernel is
\be
K^{(ND)}(s,  y, y') = \sum_{k\in\mathbb{Z}}(-1)^k\Big( K(s, y+2Lk, y') - K(s,2Lk-y, y')\Big) \,.
\lb{KNeD}
\ee
Making this heat kernel $2\pi\alpha$-periodic  and computing the trace we find
\be
\Tr K^{ND}_\alpha(s) &=& \alpha\Tr K^{ND}_{\alpha=1}(s)\nonumber \\
&&\hspace{-2.5cm} +\frac{\alpha(\alpha^{-2}-1)}{12(4\pi)^{(d-2)/2}} \frac{s\,e^{-sm^2}}{s^{d/2}}\left(A(\Sigma)\sum_{k\in\mathbb{Z}}(-1)^k e^{-\frac{L^2}{s}k^2} - \frac{1}{2}A(P)\int_0^\infty dy \sum_{k\in\mathbb{Z}}(-1)^k e^{-\frac{(y-Lk)^2}{s}} \right)\,,\quad 
\ee
and for the entanglement entropy
\be
S^{ND}_d(\Sigma) &=& S^{ND}_d(\Sigma, L)- S^{ND}_d(P)\,, \\
S^{ND}_d(\Sigma, L) &=& \frac{A(\Sigma)}{12(4\pi)^{(d-2)/2}} \int_{\epsilon^2}^\infty ds\, \frac{e^{-sm^2}}{s^{d/2}}\sum_{k\in\mathbb{Z}}(-1)^ke^{-\frac{L^2}{s}k^2} \,,\nonumber \\
S^{ND}_d(P) &=& \frac{A(P)}{24(4\pi)^{(d-2)/2}} \int_{\epsilon^2}^\infty ds\, \frac{e^{-sm^2}}{s^{d/2}}\int_{0}^{L}dy \sum_{k\in\mathbb{Z}}(-1)^ke^{-\frac{(y-Lk)^2}{s}}\,.\nonumber
\lb{SND0}
\ee 
The integration over $y$ is again given by (\ref{int}). The sum over $k$ then is vanishing,
\be
\sum_{k\in\mathbb{Z}} (-1)^k\left[(\Phi\Big(\frac{Lk}{\sqrt{s}}\Big)-\Phi\Big(\frac{L(k-1)}{\sqrt{s}}\Big)\right]=0\,,
\lb{Pk}
\ee
so that the part in the entropy (\ref{SND0}) that is due to the intersection $P$ of the entangling surface with the two boundaries is zero,
\be
S_d^{ND}(P)=0\, .
\lb{Snd}
\ee
Apparently, what happens is that the positive contribution from the Neumann boundary exactly cancels the negative contribution coming from the Dirichlet boundary
such that the total contribution is precisely zero.

For the rest of the entropy we find
\be
S^{ND}_d(\Sigma, L) &=& \frac{ A(\Sigma)}{12(4\pi)^{(d-2)/2}} \left(\int_{\epsilon^2}^\infty ds\, \frac{e^{-sm^2}}{s^{d/2}}+   {\cal S}^{ND}(L,m)      \right)\, , \\
{\cal S}^{ND}_d(L,m) &=& 
\begin{cases}\displaystyle
4\sum_{k=1}^\infty (-1)^k\Big(\frac{m}{Lk}\Big)^{\frac{d-2}{2}}K_{\frac{d-2}{2}}(2mLk)\, , &\quad m>0\,, \\
  \displaystyle \frac{4}{d-2}\frac{\Gamma(d/2)}{L^{d-2}}\sum_{k=1}^\infty\frac{(-1)^k}{k^{d-2}}\, ,& \quad m=0\,.
\end{cases}  
     \lb{kk2}
\ee
We consider some particular cases.

\bigskip

\noindent {\bf 1.} In the case of the massless field we find
\be
{\cal S}^{ND}_{d}(L, m=0)=\frac{4(2^{3-d}-1)}{(d-2)}\frac{\Gamma(\frac{d}{2})\zeta(d-2)}{L^{d-2}}\, .
\lb{11}
\ee
We note that the entropy in  dimension $d=3$ is regular now. Indeed we have in the limit
\be
\lim_{d\rightarrow 3}\,(2^{3-d}-1)\zeta(d-3)=-\ln 2\, .
\lb{11'}
\ee
This is different from what we had in the case of the boundary conditions of the same type.
In particular, it means that there is no logarithmic term in this case. So that the UV finite part in the entropy
is in fact independent of $L$,
\be
S^{\rm fin}_{d=3}(m=0)=-\frac{1}{12}\ln 2\, .
\lb{11''}
\ee

It should be noted that there have been some considerable work done in two spacetime dimensions calculating the finite size effects
in entanglement entropy. It is  curious that eq.(\ref{11''}) resembles the boundary entropy, see for instance \cite{2d}, 
in the two-dimensional case. It would be  interesting to identify  the source for this similarity. 

\medskip

\noindent {\bf 2.} In dimension $d=3$, the sum over $k$ gives
\be
\sum_{k=1}^\infty \frac{(-1)^k}{k} e^{-2mLk}=-\ln (1+e^{-2mL})\, ,
\lb{12}
\ee
and for the entropy we have
\be
{\cal S}^{ND}_{d=3}(L,m)=-2\frac{\sqrt{\pi}}{L}\ln(1+e^{-2mL}) \, .
\lb{13}
\ee
In the limit of large and small $Lm$ we  obtain
\be
{\cal S}^{ND}_{d=3}(L,m)&\simeq& -2\frac{\sqrt{\pi}}{L}e^{-2mL} \, , \  \ Lm\gg 1\nonumber \\
&\simeq& -2\frac{\sqrt{\pi}}{L}\ln2 \, , \  \  Lm\ll 1
\lb{14}
\ee

\noindent {\bf 3.} In dimension $d=5$, the sum over $k$ in (\ref{kk2}) for $m>0$ yields
\be
{\cal S}^{ND}_{d=5}(L,m)&=& \frac{\sqrt{\pi}}{L^3}\,{\rm Li}_3(-e^{-2mL}) + 2\sqrt{\pi}\frac{m}{L^2}\,{\rm Li}_2(-e^{-2mL}) \,,
\ee 
The asymptotics are given below for any $d>3$.

\noindent {\bf 4.} In dimension $d>3$ one has
\be
{\cal S}^{ND}_{d>3}(L,m)&\simeq& -2\sqrt{\pi}\,\frac{m^{\frac{d-3}{2}}}{L^{\frac{d-1}{2}}}e^{-2mL} \, , \quad\qquad Lm\gg 1\,,\\
&\simeq& -\frac{4(1-2^{3-d})}{(d-2)}\frac{\Gamma(\frac{d}{2})\zeta(d-2)}{L^{d-2}} \,,  \;\; Lm\ll 1\,.
\lb{sdd2} 
\ee

\medskip

We see that in the case of  mixed boundary conditions the UV finite part in the area law is strictly negative. This is different from what we had in the case of the same type
boundary conditions where this part in the entropy was strictly positive.

\section{Non-homogeneous entanglement: entanglement entropy density}

In almost all known explicit calculations of entanglement entropy, the entangling surface (plane, sphere, cylinder) has a large group of symmetry (a combination of rotations and translations). This symmetry indicates that all points on the surface are equivalent in the sense that neither of them is in a preferred position. 
When the entanglement entropy is computed for such a surface, the symmetry tells us that the entanglement across the surface is homogeneous, i.e. it
is the same for all points on the surface. That is why, to leading order, the entanglement entropy is  simply proportional to the area of the surface.

In the cases where the entangling surface intersects the boundary, the situation changes. The symmetry is now broken by the presence of the boundary.
This is clearly the case in the examples considered above where the plane surface intersects the plane boundaries. Thus in these examples we may expect that the
entanglement is not homogeneous along the surface, and that the points close to the boundary are in a certain sense more preferred than those lying far from the boundary.
In order to describe quantitatively this non-homogeneity, we introduce the density of entanglement entropy which characterizes the {\it local   entanglement} along the surface.

In all cases considered in this paper the entanglement entropy is obtained by taking two integrations, one with respect to the proper time $s$ and the second
is with respect to the coordinates  $(y, x^i, \ i=1,..,d-3)$, where $y$ is orthogonal to the boundary and $x^i, \ i=1,..,d-3$ are the other coordinates on $\Sigma$,
\be
S=\int_{\epsilon^2}^\infty ds \int d^{d-3}x\int_0^\infty dy ~S(s, y,x)\, ,
\lb{8.1}
\ee
Interchanging the order  of integration in (\ref{8.1}) we have
\be
S=\int_{\epsilon'}^\infty dy \int d^{d-3}x~S(x,y)\, , \ \ S(x,y)=\int_0^\infty ds~S(s,x, y)\, .
\lb{8.2}
\ee
The quantity $S(x,y)$ we shall call the entanglement entropy density. In all examples considered in the paper, $S(x,y)$ is function of variable $y$ only, so that the entanglement is homogeneous in the directions orthogonal to $y$. We notice that after the interchange, the integration over $s$ for any finite $y$ may be 
well defined so that no regularization in lower limit would be necessary. Instead the integration over $y$ of the entropy density may lead to some divergences for small values of
$y$ such that a new regularization, with a  regularization parameter $\epsilon'$, would be necessary. 
In all cases considered below the entropy density has two contributions: a constant (homogeneous) contribution $S_h$ and a non-homogeneous contribution $\texttt{S}(y)$,
\be
S(s,x,y)=S_h+\texttt{S}(y)\, .
\lb{8.3}
\ee
The homogeneous piece, once integrated,  will give rise to the area law in the entropy. 
In this section, we are mainly interested in the non-homogeneous part. This piece  will quantitatively characterize  how the quantum entanglement  changes with $y$.

\bigskip

\subsection{Single plane boundary: Neumann(Dirichlet) boundary condition}

The entanglement entropy in this case is given by (\ref{SND}). We find 
\be
S(s,x,y)=S_h\pm\frac{1}{12(4\pi)^{\frac{d-2}{2}}}\frac{e^{-sm^2}}{s^{d/2}}e^{-y^2/s}\, .
\lb{8.4}
\ee
The homogeneous piece $S_h$ will produce the term $S_d(\Sigma)$ in the entropy while the second term in (\ref{8.4}) will give rise to $S_d(P)$ if integrated over $(y,x)$. 
Interchanging the integration over $s$ and $(y,x)$ we find for the non-homogeneous piece in  the entropy density,
\be
\texttt{S}(y)=
\begin{cases}\displaystyle
\pm \frac{1}{6(4\pi)^{\frac{d-2}{2}}}\Big(\frac{m}{y}\Big)^{\frac{d-2}{2}}K_{\frac{d-2}{2}}(2my)\, , &\quad m>0\,, \vspace{7pt}\\
  \displaystyle \pm \frac{\Gamma(d/2)}{6(d-2)(4\pi)^{\frac{d-2}{2}}}\frac{1}{y^{d-2}}\, ,& \quad m=0\,.
\end{cases} 
\lb{8.5}
\ee
We see that this density decays fast with the distance from the boundary and becomes divergent when one approaches the boundary. This behavior indicates that
the local  quantum entanglement is maximal near the boundary and falls off  with the distance from the boundary.

\subsection{Single plane boundary: Robin boundary condition}

In this case we find
\be
S^{(h)}(s,x,y)=S_h+\frac{1}{12(4\pi)^{\frac{d-2}{2}}}\frac{e^{-sm^2}}{s^{d/2}}\left(e^{-y^2/s}-2he^{2hy}\int_{2y}^\infty d\sigma e^{-h\sigma}e^{-\frac{\sigma^2}{4s}}\right) \, .
\lb{8.6}
\ee
The integration over $s$ then will give us
\be
\texttt{S}^{(h)}(y)=\frac{m^{\frac{d-2}{2}}}{6(4\pi)^{\frac{d-2}{2}}} \left(y^{\frac{2-d}{2}}K_{\frac{d-2}{2}}(2my)-2^{d/2}he^{2hy}\int_{2y}^\infty d\sigma\, \sigma^{\frac{2-d}{2}}e^{-h\sigma}K_{\frac{d-2}{2}}(m\sigma)\right)\, .
\lb{8.61}
\ee
In the massless case ($m=0$)  the integration over $\sigma$ is expressed in terms of the incomplete Gamma function
and one finds
\be
\texttt{S}^{(h)}(y)=\frac{\Gamma(d/2)}{6(d-2)(4\pi)^{\frac{d-2}{2}}} \left(\frac{1}{y^{d-2}}-2^{d-1}h^{d-2}e^{2hy}\,\Gamma(3-d, 2hy)\right)\, .
\lb{8.7}
\ee
\begin{figure}[ht]
\begin{center}
\begin{tikzpicture}[scale=0.95]
\draw[black,thick,dashed] plot[smooth] coordinates {(-3.75,3.99) (-2.5,0.3) (4.4,-0.22)};
\draw[black,thick,dashed] plot[smooth] coordinates {(-3.65,-3.6) (-2.4,-0.78)  (4.4,-0.22)};
\draw [thick] [->] (-4.35,-3.6) --++ (0,7.75) node [above] {$\texttt{S}^{(h)}(y)$};
\draw [thick] [->] (-4.35,-0.23) --++(9,0) node [right] {$y$};
\node at (0,0) {\includegraphics[height=7.6cm]{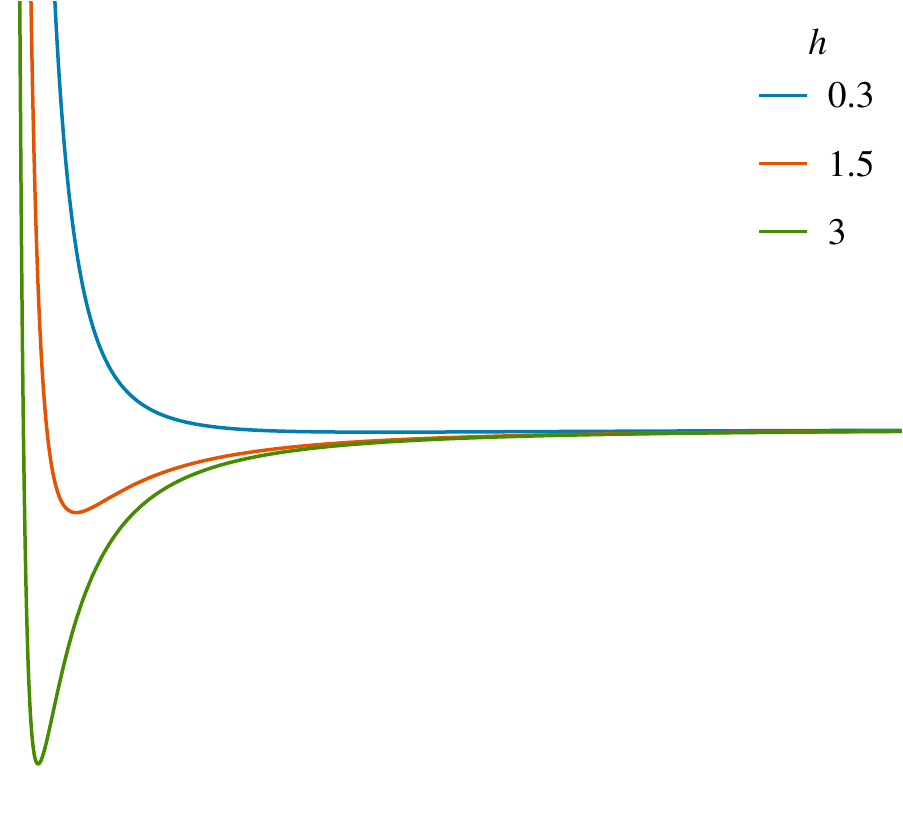}};
\end{tikzpicture}
\end{center}
\vspace{-1cm}
\caption{Entropy density for single plane Robin BC in $d=4$ dimensions. Dashed plots correspond to Neumann (up) and Dirichlet (down) boundary condition.}
\end{figure}
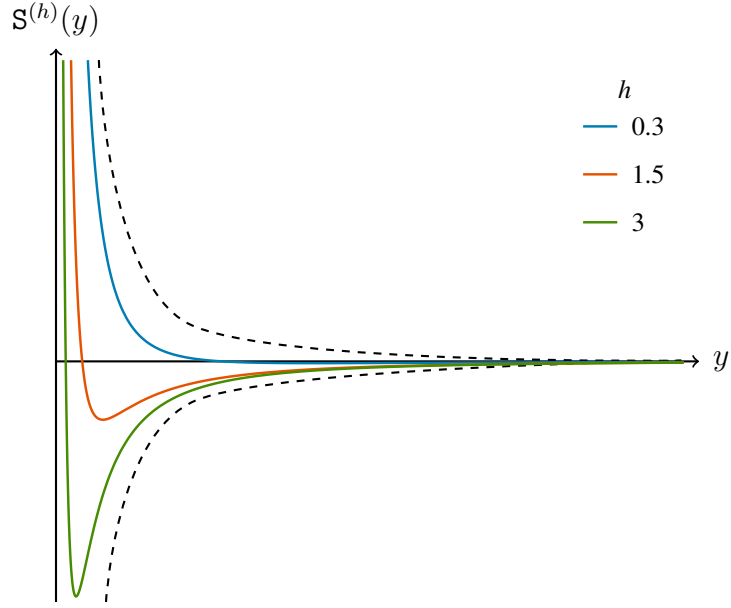
For large values of $y$ this function becomes negative and approaches the Dirichlet function (\ref{8.5}) while for small values of $y$ it is positive and is approximated by the Neumann function
(\ref{8.5}). 
The separation between the two regimes is governed by $1/h$ scale so that in the limit of large $h$ the Dirichlet region becomes dominating
while for small $h$ the Neumann region dominates.  Similar behavior is expected for non-vanishing mass $m$.

\subsection{Two plane boundaries}
For simplicity we shall consider the massless case, $m=0$.
First we consider the case of the same type (Neumann or Dirichlet) boundary conditions at $y=0$ and $y=L$.
In this case
\be
S^{NN(DD)}(s,x,y)=S_h\pm \frac{1}{12(4\pi)^{\frac{d-2}{2}}}s^{-d/2}\sum_{k\in\mathbb{Z}}e^{-(y-Lk)^2/s}\, .
\lb{8.8}
\ee
The integration over $s$ for the non-homogeneous part will give us
\be
\texttt{S}^{NN(DD)}_d(y)=\pm\frac{\Gamma(d/2)}{6(d-2)(4\pi)^{\frac{d-2}{2}}}\sum_{k\in\mathbb{Z}} \frac{1}{|y-Lk|^{d-2}}\, .
\lb{8.9}
\ee
For even $d>2$ the sum results in elementary functions. Here are some examples:
\be
 \sum_{k\in\mathbb{Z}} \frac{1}{|y-Lk|^{2}} &=& \frac{\pi^2}{L^2\sin^2(\frac{\pi y}{L})}\, , \ \ \ \ \  \ \ \ \ \ \  d=4 \nonumber \\
 \sum_{k\in\mathbb{Z}} \frac{1}{|y-Lk|^{4}}  &=& \frac{\pi^4}{3L^4}\frac{2\cos^2(\frac{\pi y}{L})+1}{\sin^4(\frac{\pi y}{L})}\, , \ \  d=6
 \lb{8.10}
 \ee
For any $d>3$ we find
\be
 \sum_{k\in\mathbb{Z}} \frac{1}{|y-Lk|^{d-2}} &=& \frac{1}{y^{d-2}} + \frac{1}{L^{d-2}}\Big( \zeta(d-2,1+y/L)+ \zeta(d-2,1-y/L) \Big)\,. 
\ee

\medskip

In the case of mixed boundary conditions we have
\be
S^{ND}(s,x,y)=S_h+ \frac{1}{12(4\pi)^{\frac{d-2}{2}}}s^{-d/2}\sum_{k\in\mathbb{Z}}(-1)^ke^{-(y-Lk)^2/s}\, .
\lb{8.11}
\ee  
The integration over $s$ for the non-homogeneous part will give us
\be
\texttt{S}^{ND}_d(y)=\frac{\Gamma(d/2)}{6(d-2)(4\pi)^{\frac{d-2}{2}}}\sum_{k\in\mathbb{Z}} \frac{(-1)^k}{|y-Lk|^{d-2}}\, .
\lb{8.12}
\ee
\begin{figure}[ht]
\begin{center}
\begin{tikzpicture}[scale=0.95]
\hspace{-0.1cm}
\draw [thick] [->] (-3.9,-2.67) --++ (0,5.5) node [above] {$\texttt{S}^{NN}(y)$};
\draw [thick] (3.9,-2.67) --++ (0,5.4);
\draw [thick] (-3.9,-2.65) --++(7.8,0) node [below] at (0,-2.7) {$L$};
\node at (0,0) {\includegraphics[height=6.95cm]{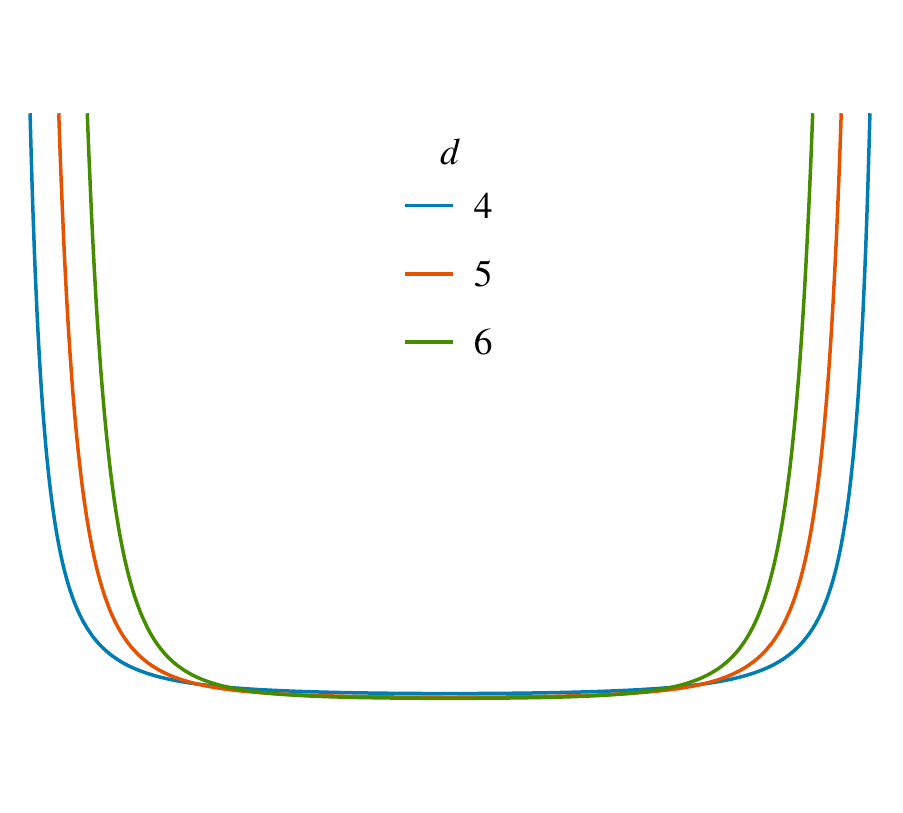}};
\draw [thick] [->] (5.2,-3.5) --++ (0,7.1) node [above] {$\texttt{S}^{ND}(y)$};
\draw [thick] (12.8,-3.5) --++ (0,7.);
\draw [thick] (5.2,0.0) --++(7.6,0) node [below] at (9,-0.05) {$L$};
\node at (9,0) {\includegraphics[height=6.5cm]{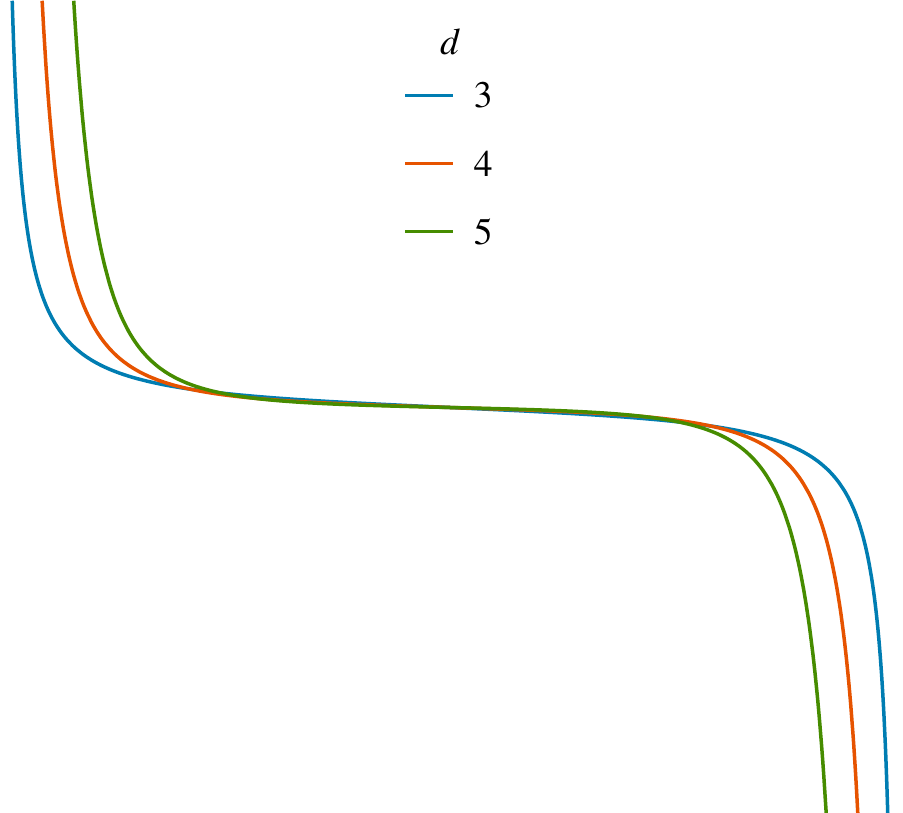}};
\end{tikzpicture}
\end{center}
\vspace{-0.8cm}
\caption{Entropy densities for two planes boundaries. Neumann-Neumann (left) for $d=4,\,5\,, 6$ dimensions and Neumann-Dirichlet (right) for $d=3,\,4\,, 5$ dimensions.}
\end{figure}
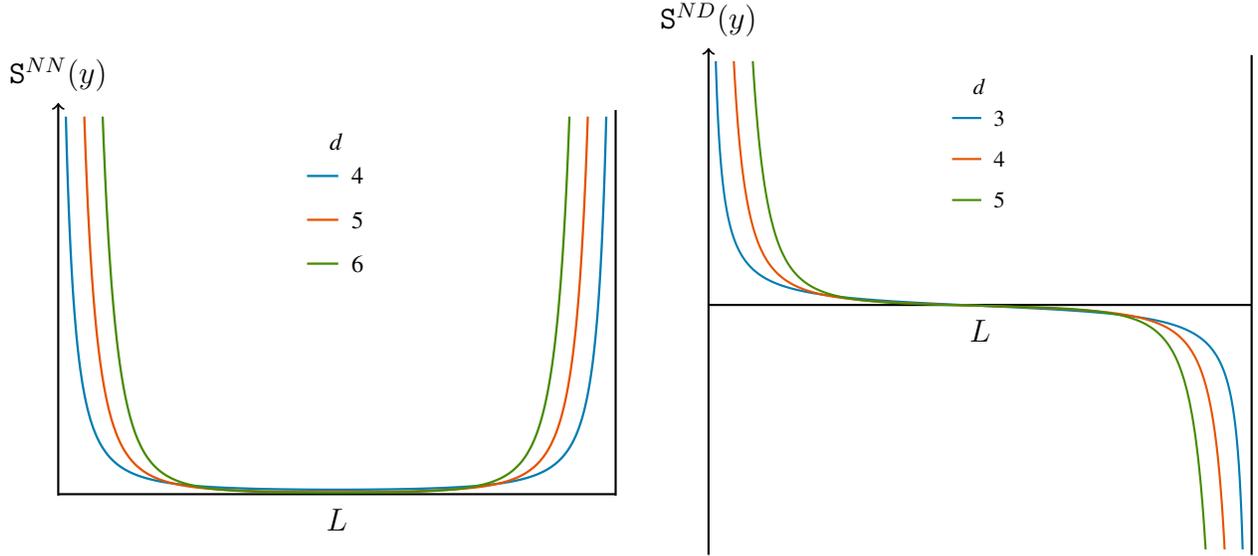
By separating the sum into two parts, for even and odd integers, we find a relation to the entropy in the case of NN boundary conditions,
\be
\texttt{S}^{ND}_d(y, L)=\texttt{S}^{NN}_d(y, 2L)-\texttt{S}^{NN}_d(y-L, 2L)\, .
\lb{8.13}
\ee
In particular, in dimension $d=4$ we have
\be
\texttt{S}^{ND}_{d=4}(y, L)=\frac{\pi}{192L^2}\left( \frac{1}{\sin^2(\frac{\pi y}{2L})}- \frac{1}{\sin^2(\frac{\pi (y-L)}{2L})}\right)\, .
\lb{8.14}
\ee
The relation (\ref{8.13}) shows that closer to the boundary at $y=0$ the non-homogeneous part in the entropy approaches the one of the Neumann boundary while closer to 
boundary at $y=L$ it approaches the entropy of the Dirichlet case. Clearly, the function (\ref{8.13}) flips the sign under the reflection ($y\rightarrow L-y$) with respect to the
plane at $y=L/2$.  This explains why the integrated quantity vanishes as we have shown earlier in the paper  (see  (\ref{Snd})).

\bigskip

All these examples show that the quantum entanglement is stronger closer to the boundary. Taking into account the presence of the constant, homogeneous, piece in the entropy density
and the sign in the entropy density we conclude that it is near the Neumann boundary where the entanglement is maximal. The monotonicity of the entanglement entropy under the flow in the boundary coupling then  can be interpreted as the monotonic decreasing in the local quantum entanglement.

\section{Conclusions}
In this paper we have presented several explicit calculations of entanglement entropy in the presence of boundaries.
In the cases  where the entangling surface intersects the boundary at a  co-dimension three surface  $P$ there appear new terms in the entropy.
Those terms are defined at  the surface $P$ and they depend on the type of the boundary condition imposed.
We have considered the Neumann, Dirichlet and Robin type conditions. The latter is the most general one. By changing the parameter
that appears in the Robin condition one interpolates between the Neumann and the Dirichlet conditions.  Among other findings we
prove that the entanglement entropy is monotonically decreasing in the course of  this interpolation.  This may have some applications in the analysis
of the RG flow in the presence of boundaries.

The situations where the entangling surface intersects the boundaries provide us with important examples in which the quantum entanglement
is not homogeneous along the surface. We demonstrate on a number of such examples that the entanglement is stronger closer to the
boundaries.  It would be interesting to verify this prediction in an experiment.

The calculations considered in this paper should give us some intuition on what may happen in more general situations, when the spacetime is  non-flat and
the boundary as well as the entangling surface  are curved.  Since many factors come into play, the entropy in such a general situation may be rather 
complicated and difficult to calculate. We, however, anticipate that our findings: the locality of the contribution due to intersection $P$,
decay of entropy with  the size of the system, the monotonicity of the entropy in the flow between the Dirichlet and Neumann phases,
are universal and should be present in these more general situations.

Finally,  we note that all physical systems around us are confined to some boundaries. Therefore, if the entanglement  entropy, or its variation,  will be ever measured  in an experiment the role of the boundaries should be important if not decisive.  This certainly motivates the necessity of the further study of boundary effects in entanglement entropy.


\end{document}